\documentclass[sigconf]{acmart}

\usepackage{booktabs} 

\setcopyright{none}

\acmConference[Preprint]{September}{Conf.}{2017} 
\acmYear{2017}
\copyrightyear{2017}

\acmArticle{4}
\acmPrice{0.00}

\begin{document}
\title{Microservices: Granularity vs. Performance}

\author{Dharmendra Shadija}
\orcid{1234-5678-9012}
\affiliation{%
  \institution{Sheffield Hallam University}
  \streetaddress{P.O. Box 1212}
  \city{Sheffield} 
  \state{United Kingdom} 
  \postcode{43017-6221}
}
\email{d.shadija@shu.ac.uk}

\author{Mo Rezai}
\orcid{1234-5678-9012}
\affiliation{%
  \institution{Sheffield Hallam University}
  \city{Sheffield} 
  \state{United Kingdom} 
}
\email{m.j.rezai@shu.ac.uk}

\author{Richard Hill}
\orcid{1234-5678-9012}
\affiliation{%
  \institution{University of Huddersfield}
  \streetaddress{Queensgate}
  \city{Huddersfield} 
  \state{United Kingdom} 
}
\email{r.hill@hud.ac.uk}

\renewcommand{\shortauthors}{D. Shadija et al.}

\begin{abstract}
Microservice Architectures (MA) have the potential to increase the agility of software development. In an era where businesses require software applications to evolve to support software emerging requirements, particularly for Internet of Things (IoT) applications, we examine the issue of microservice granularity and explore its effect upon application latency. Two approaches to microservice deployment are simulated; the first with microservices in a single container, and the second with microservices partitioned across separate containers. We observed a neglibible increase in service latency for the multiple container deployment over a single container.

\end{abstract}

%
%



\maketitle



\section{Introduction}
Whilst the availability of utility clouds reduces the costs and associated responsibilities for physical infrastructure for business IT functions, it also presents new opportunities for application development teams to exploit. Providers of utility cloud computing endeavour to deliver seamless computing and storage services, that are elastic in nature, meaning that software development for cloud computing can focus more on flexibility, reuse and improved Quality of Service (QoS).

Service Oriented Architecture (SOA, and subsequently ``web services''\cite{berners-lee2009,alonso2004}), is a natural fit for ``everything-as-a-service'', but it is also practical to decompose software applications into discrete services as it can help bridge the comprehension gap between users requirements and design specifications, whilst also improving software design by moving away from more inflexible, monolithic architectures\cite{walker2007,hasselbring2016}.

As service orientation thinking matures, there is now the construct of Microservice Architecures (MSA)\cite{lewis2014}, which have gained popularity with software development teams who have a need to be able to provide applications that can scale in response to emerging requirements\cite{thones2015}. An MSA consists of discrete services that are interconnected to deliver a workflow\cite{dragoni2017a}\cite{dragoni2017}.

In general, an MSA can be thought of has containing services that satisfy more focused areas of an applications functionalities, though this is not always the case. Further discussion of this can be found in Shadija et al\cite{shadija2017}.

The need to consider applications that can support future expansion in functionality is a logical progression as computational and communication capabilities become embedded in more devices\cite{Bessis2013}. Specifically, the Internet of Things (IoT) presents many new and in some cases unforeseen ways for users to interact with systems, as well as systems being able to sense their environment in a multitude of different scenarios through IoT devices\cite{ikram2015}.

This article explores one particular aspect of MSA - service granularity - as this has considerable potential to have an impact upon application latency\cite{bryant2016}. We examine the deployment of an enterprise software application using two MS architectures, and suggest some indicative guidelines for application architects to consider when designing or migrating to cloud-based applications that utilise microservices. We then discuss the findings in the context of IoT application architectures.

The article is organised as follows. First we introduce the key concepts and qualitative issues surrounding MS granularity. We then apply MSA principles to the cloud deployment of a university admissions system and simulate service invocation response times. Finally, we discuss MSA deployment in the context of IoT and identify some open questions for the research community.
\section{Granularity}
Microservices can be declared with varying levels of capability, and the size of this functionality is typically referred to as its granularity, that is, the functional complexity coded in a service or number of use cases implemented by a microservice\cite{newman2017}.

Since microservices are discrete and must be composed into greater functional entities to support business workflows, it follows that message passing between microservices (as a result of method invocation) increases as the microservices become finer-grained.

The `building-block' approach to service composition is attractive from an architectural perspective; arguments for service re-use can be made, and the gap between application design and the user requirements documentation can be reduced. However, the increase in communication between services (manifesting as out-of-process calls and the number of service calls made) also increases the response time of an application, particularly when many small increases in latency are compounded together\cite{xu2008}.

Achieving an optimum level of granularity is therefore of interest to application developers who want to explore MSA for deployment, and the key factors that contribute to this can be summarised as follows:
\begin{itemize}
\item \textit{Driven by business need or capability.} The needs of a business may be changing rapidly and demanding new functionality from an application. This growth may not be manageable within the existing application architecture and therefore a granular approach is adopted. It is typical for application developers to use the functionality itself to set the scope that determines the size of a microservice.

\item \textit{Size of application.} For smaller applications the level of granularity could be fine-grained. For enterprise (larger) sized applications, the granularity is likely to be at a higher level (coarser) with each microservice built up from smaller microservices. However, as we discuss later, for smaller applications there may stiil need to be an aggregation of services to facilitate simpler communication and reduced latency over IoT network connections.

\item \textit{Size of development team.} The number of developers in a team, together with their skills capability should be considered. Conway (\url{http://www.melconway.com/Home/Conways_Law.html}) says ``organizations which design systems ... are constrained to produce designs which are copies of the communication structures of these organizations''. In the context of MSA, but more specifically Domain Driven Design \cite{evans2003}, the degree of success of functional decomposition, and its subsequent implementation as a successful service, is dependent upon the organisational structure of the development teams.

\item \textit{Database design.} The design of a database may have an impact on granularity. For example, in a retail scenario if there is a \textit{product} service and an \textit{order} service, the functional decomposition is likely to have led to the implementation of  separate data repositories for each service. Any association of the data between the databases will be implemented at code level, leading to more coarse-grained microservices.

\item \textit{Reuse.} MSA promotion of reuse in the architecture is a concern for enterprise applications. If the services are fine-grained then reuse is possible but there is the additional overhead of wiring the services together. If the services are too coarse grained then it is difficult to reuse the services.
\end{itemize}
One of the main concerns for application developers to consider is that of user experience, which is influenced by the perceived performance of an application. Developers seek to minimise latency within an application as far as is practicable, in order to maximise software responsiveness\cite{newman2017,viennot2015}.

\begin{figure}[!t]
  \caption{Method invocation between microservices.}
  \centering
    \includegraphics[width=0.25\textwidth]{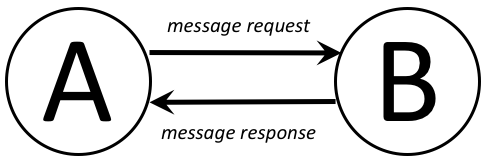}
    \label{fig:methodinvocation}
\end{figure}

It follows that whilst architectural concerns may lead designers towards finer-grained functional decomposition\cite{evans2015,evans2003}, any potential increase in the number of methods invoked, either within a container or between physical servers via a network, will have an increased contribution towards latency, particularly when virtualised. Containers are one means of addressing the latency to some extent when the microservices exist in cloud environments\cite{kubernetes2017,docker2017}. Figure\ref{fig:methodinvocation} illustrates the message request and response between two microservices, \textit{A} and \textit{B}.
The overall latency is determined by a number of factors, including the following:
\begin{itemize}
\item Number of calls;
\item Network latency and availability;
\item Availability of microservice;
\item Processing time;
\item Variability in demand/load;
\end{itemize}
While considering latency there are two other factors to consider. First, the criticality of the service being called and second, the number of times it is called. For example, although calling a ``qualification mapping'' service is important, it is not critical to the function of the system. If it is not working the system can still accept new applications.
%


An additional factor to be considered is whether the application under consideration is a new application or an existing application that is being migrated to MSA. For a new application, database design will have less influence as there will not be a database in place already. For an existing application, the organisation and design of existing data and their structures is an important factor to consider. 


This article uses an evolutionary case study to propose guiding principles for granularity within MSA. As MSA has an affinity with systematic growth in an application, we have chosen a case study that depicts expansion in functionality and user load over time, making it an ideal candidate for MSA. 
\section{Case Study: University Admissions System}\label{sec:case_study1}
UK universities each administer their own applications to receive and process applications from potential students. In the UK, the overall process is governed by a Central Admissions Authority (CAA), the Universities and Colleges Admissions Service (UCAS), with each higher education institution deploying local systems to interface with the national UCAS system.

CAA systems are examples of applications that traditionally had stable requirements in terms of specification, and it was rare that additional functionality needed to be added. However, changes in the UK national policies for students obtaining funding for university study have led to a shift in behaviour in relation to how applicants choose their institution. 

Such a change has led to universities seeking to reach out to their applicants in new ways, requiring closer integration between previously disparate systems such as marketing, as well as developing new capabilities to engage applicants.

Additionally, university admissions systems are examples of large scale applications that are now being migrated to clouds, with an associated degree of effort being invested into the translation of traditional monolithic structures into service-oriented models.

Therefore, changes in the higher education environment are placing demands upon the providers of CAA systems. 

The proliferation of mobile devices being used as interfaces to larger systems is leading to a desire to enable tighter integration of a greater range of devices, including RFID personal identification cards, location sensing and mobility tracking technology. This potential rich environment of technology-facilitated interaction lends itself, at least conceptually, to an MSA.

As such, we consider the case whereby an existing CAA admissions system is to be re-engineered using an MSA. We also illustrate the expansion and growth in requirements of the application over time, summarised by the following four stages.
\subsection{Stage One}
Intial motivations for the application are:
\begin{enumerate}
\item Accurately recording applicant information.
\item Recording the offer received from each university.
\end{enumerate}
\subsubsection{System workflow}
Each applicant completes a paper form which is posted to the CAA together with any required documentation. An Admissions Officer (AO) at the CAA enters details from the form into their system. Forms are then printed by the AO and posted to the relevant universities. Universities assess individual applications and make offers. Universities communicate offers directly to students.
\begin{table*}
  \caption{REST API call URLs}
  \label{tab:restcalls}
  \begin{tabular}{llll}
    \toprule
    Method&URL&Description&Database action \\
    \midrule
    GET    & /CAA & Entry point to CAA Application & READ \\
    GET    &/CAA/Application/&Reads application(s) from Database & READ \\
    GET    &/CAA/Application/\{applicationId\} &Reads application from Database based on applicationId & READ\\
    POST   &/CAA/Application/\{applicationId\}& Updates an application in the database& UPDATE \\
    PUT    &/CAA/Application/ & Adds a new application into the database. & INSERT \\
    DELETE &/CAA/Application/\{applicationId\}&Deletes an application from database & DELETE\\
    GET    &/CAA/Application/PrintApplication & Prints application for delivery to university & READ \\
    POST   &/CAA/Application/ShipApplication & Delivers application to university & POST \\       
  \bottomrule
\end{tabular}
\end{table*}
%
%
\subsection{Stage Two}
Students search for course information on the CAA system. Currently this information is supplied by each university to the CAA on an annual basis.  One of the flaws of this system is that information could be out of date due to new courses being introduced or some courses being deleted from the prospectus. 
\begin{itemize}
\item In addition to functionality from Stage One, the CAA system should also display up to date course information.
\end{itemize}
\subsubsection{System workflow}
In this case, the application workflow from a student perspective remains the same. There is a change in how course information is displayed to potential students; the system now accesses course information in real time using REST services. 
\subsection{Stage Three}
Due to the success of the CAA application it was decided that college admissions should also be undertaken using the CAA system.
\subsubsection{System workflow}
Again, the application workflow remains the same from the perspective of a student, with the addition that they now have the option of applying for college (Further Education) admission as well as university (Higher Education). 
\subsection{Stage Four} 
The CAA System will now process applications from international (overseas) students. This also integrates functionality from the qualification mapping services in external organisations, during the application handling process. 
The REST API calls to achieve the above functionality are summarised in Table \ref{tab:restcalls}. An overview of the MSA deployed onto one container is illustrated in Figure \ref{fig:ucassinglecontainer}.

%

%
\section{Simulation}
\begin{figure*}[!tb]
  \caption{Application architecture with microservices deployed into a single container.}
  \centering
    \includegraphics[width=0.79\textwidth]{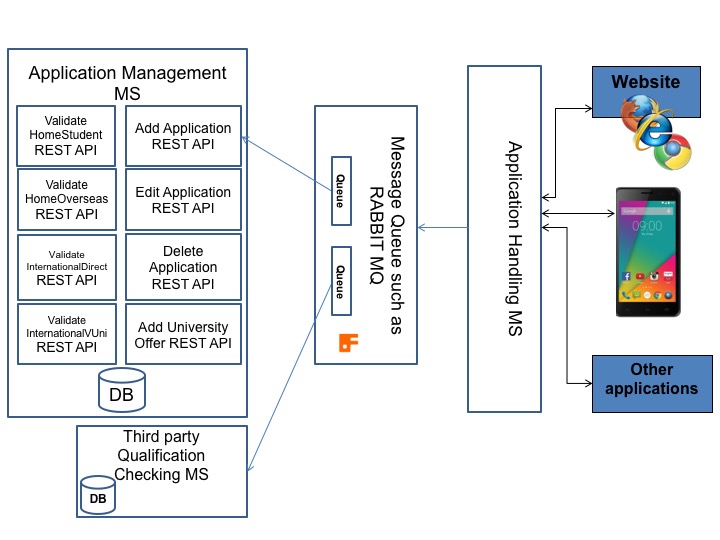}
    \label{fig:ucassinglecontainer}
\end{figure*}

\begin{figure*}[!tb]
  \caption{Application architecture with microservices deployed into two containers.}
  \centering
    \includegraphics[width=0.79\textwidth]{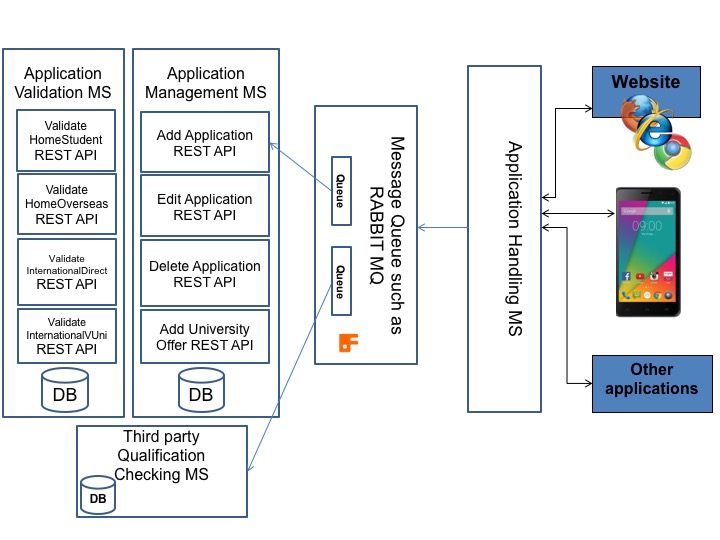}
    \label{fig:ucasseparatecontainers}
\end{figure*}

From the application scenario described in section \ref{sec:case_study1}, we have simulated two MSA deployments. The first uses a single container for all of the microservices as per Figure \ref{fig:ucassinglecontainer}.
Figure \ref{fig:ucasseparatecontainers} illustrates the partitioning of microservices onto two containers, thereby introducing method invocation across containers.
\subsection{Test environment}
For the simulation, we deployed Microsoft ASP.Net REST APIs programmed using C\# onto a Microsoft IIS v7 webserver. The web application was also hosted on an IIS webserver. Activity monitoring was provided by the Dynatrace (https://www.dynatrace.com/) tool.
\subsection{Results}
\begin{figure}[!htb]
  \caption{Round-trip service requests/responses and latency for single and dual container deployment.}
  \centering
    \includegraphics[width=0.47\textwidth]{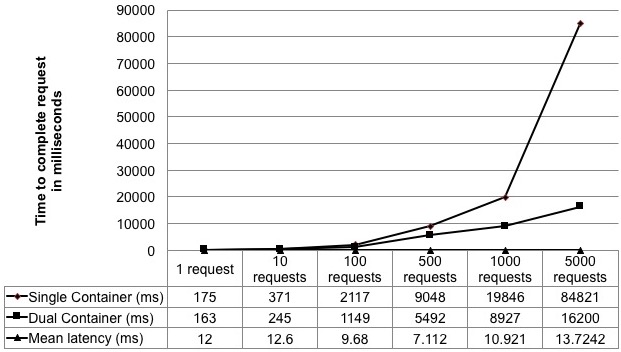}
    \label{fig:results_graph}
\end{figure}

The results are summarised in Figure \ref{fig:results_graph}.
In the first case, all of the microservices identified were implemented and hosted on the same server, and requests of varying volume were made through the web application. The roundtrip time of service invocation was recorded using Dynatrace. This was repeated for the dual container implementation, where the microservices were distributed across the web servers.

Figure \ref{fig:results_graph} quantifies the latency for each case, showing how latency increases for a given volume of requests from the user interface. The graph shows the total time taken for a request originating from the browser to a response being displayed in the browser.
%
%
%
%
In the second case services were implemented on separate servers. Since a network component had been inserted between the services an increase in latency was expected.

However the relative difference between each deployment remains small at less than 1\%, though this is for a single workflow with no dependencies upon other supplying services. Since MSAs promote re-use, nested dependencies have the potential to introduce significant delays even if the origins of the latency are relatively small.
%
%
%
%

\section{Discussion}
Finer-grained microservices result in a potential increase in the number of in-process method invocations. This is tolerable when the microservices lie within the same container, as the message passing is rapid and the probability that the service request is completed is high and wholly dependent upon the application in the container.

Once the messages broach the boundaries of containers to make requests to external microservices, additional vulnerabilities threaten the successful completion of the request, not least the variables introduced by a network connection.

As such, the issue of performance is not wholly restricted by latency through increased network traffic, but it is also influenced by additional risks from external communication mechanisms.

An application designer who does not have to consider distributed applications can then adopt the MSA approach within one container, or at least use containers to scale an application but have those containers hosted on the same cloud.

However, this is not the case for applications that require the integration of IoT devices, whereby it is expected that there will be a distribution of services across entities, and that there will be at least two containers on at least two hosts.

This suggests that the issue of microservice granularity has to be treated differently depending upon the future requirements that are envisaged, which is the underpinning challenge and therefore the motivation for this work. Microservices that co-exist in the same container can benefit from finer-grained deployments, whereas microservices distributed across networks have a more resilient architecture when they are exposed as courser-grained services.
\subsection{Implications for IoT applications}
Approaches to IT delivery such as cloud computing have enabled application development to be somewhat simplified, which in turn has facilitated more complex solutions to challenging business problems. The Internet of Things does provide far more challenging complexity to deal with.

The constraints of device capability are rapidly dwindling as the miniaturisation of hardware continues to reduce costs and increase capabilities. One such example is edge computing, where computational power is located at the edge of a network as opposed to being centralised as per the client-server or n-tier architectures.

Edge computing utilises \textit{cloudlets} to simplify collections of constrained hardware (such as embedded systems, FPGA, etc.) so that applications can be deployed in a platform and technology agnostic fashion \cite{hill2017}.

\begin{figure}[!t]
  \caption{Using cloudlets to simplify streaming analytics at the edge of a network.}
  \centering
    \includegraphics[width=0.50\textwidth]{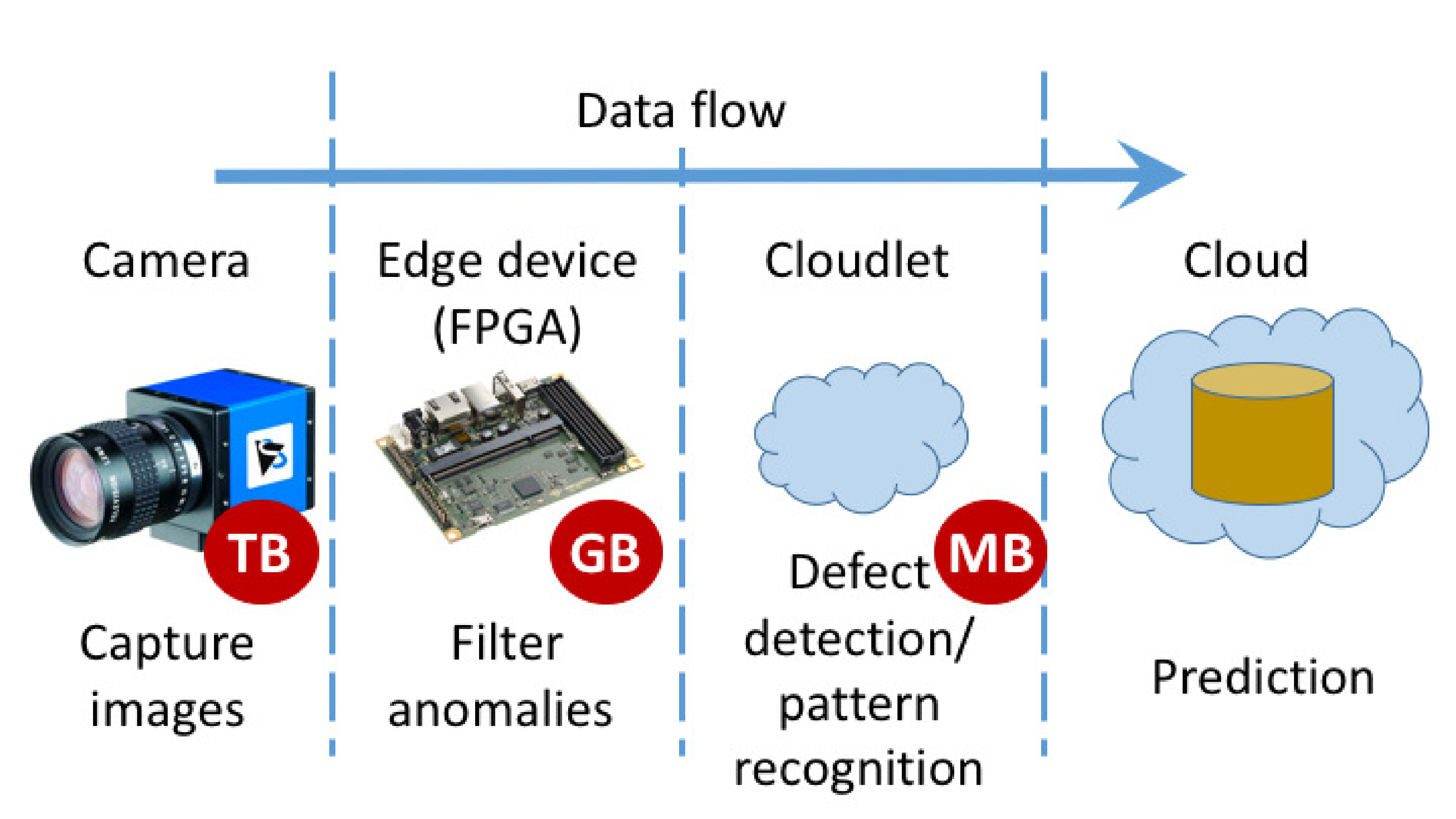}
    \label{fig:itacasestudy}
\end{figure}

Figure \ref{fig:itacasestudy} describes an edge computing architecture whereby digital camera images are pre-processed on an edge device, before being mined for patterns on a cloudlet. The cloudlet is one or more devices local to the camera, and therefore in close proximity to the edge of the network.

\begin{figure}[!t]
  \caption{An IoT architecture for the delivery of community healthcare services.}
  \centering
    \includegraphics[width=0.50\textwidth]{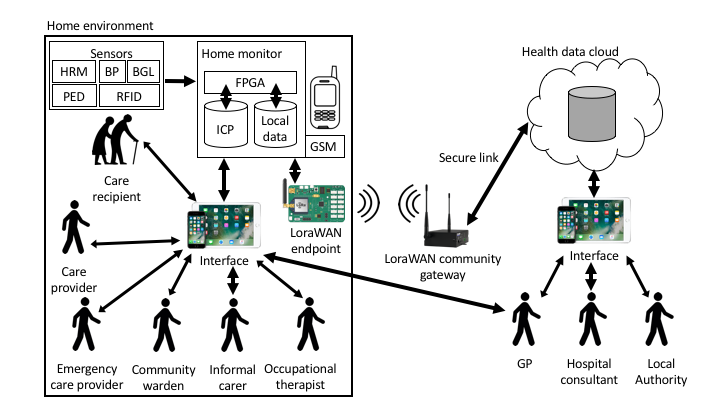}
    \label{fig:inca2architecture}
\end{figure}

Microservice messaging deployed within the cloudlet will be a combination of in-container and container-to-container calls, as well as communicating with services hosted in the destination cloud. ZeroMQ is one example of a protocol for inter-service communication\cite{zeromq2017}.
Similar examples exist in other domains such as the delivery of community healthcare services\cite{beer2003d,Hill2005d,hill2017a}, where there is the added complexity of restricted bandwidth in Low Power Wide Area Networks (LPWAN), used as an effective way of ensuring reliable network coverage that is independent of telecommunications or WiFi networks (see Figure \ref{fig:inca2architecture}).
Data collected via IoT devices is analysed locally within the home environment, providing a fast response for visualisation, whilst also maintaining patient privacy. Analytics functions then produce insight that is stored in a central remote cloud, upon which intelligence functions can be performed upon the collective dataset\cite{aqrabi2012a} of the community. These domains are challenging to model so that the interactions can be authentically captured\cite{beer2003}

MSA seems to be particularly suited to these emerging situations as it is feasible that services may be decomposed and distributed in one setting, and then bundled together in fewer (or single) containers in another scenario.

This implies that the deployment domain might influence the orchestration of the \textit{granularity} of microservices; this is an interesting open challenge for the research community, particularly with regard to the autonomous orchestration, packaging and deployment of microservices as a result of sensing the local IoT environment.
\section{Conclusions}
%
Microservices appear to enable a more detailed `finer-grained' approach to service declaration, and as a consequence can permit greater reuse of functionality when they co-exist within a container, or alongside separate containers on the same host. Whilst this traditionally would have been a server, this also holds true for cloud based environments where the platform is abstracted away from the hardware via virtualisation. This level of granularity is much finer than that experienced with more established web services.

However, once the requirement for service invocation requires a call to a container or host via a network link, there is a reduction in performance both as a result of network data transfer rates, as well as the potential for a communication link to fail.

The decision process for microservice granularity is therefore influenced by a number of environmental factors, irrespective of the potential to rely purely upon functional decomposition to specify microservices at the ``correct'' size. In heterogeneous environments such as clouds, abstraction from the disparate hardware is provided by the platform. As yet, this concept of abstraction is not apparent in IoT scenarios, and therefore, with the implied reliance upon network links (and most likely those links will have limited bandwidth), granularity must be considered from the standpoint of its eventual impact upon application performance when deploying MSA.


\bibliographystyle{ACM-Reference-Format}
\bibliography{rh}
\end{document}